\documentclass[twocolumn,showpacs,preprintnumbers,prd,superscriptaddress,nofootinbib]{revtex4}
\usepackage{comment}
\usepackage{graphicx}
\usepackage{amsmath, mathrsfs}
\usepackage{float}
\usepackage{subfigure}
\usepackage[usenames]{color}
\usepackage{amssymb}
\usepackage{bbm}
\usepackage{array}    
\usepackage{csquotes}
\usepackage{color}
\usepackage{psfrag}
\usepackage{graphicx}
\usepackage{caption,subcaption}
\usepackage{marginnote}
\usepackage[utf8]{inputenc}
\usepackage{tikz}
\usetikzlibrary{positioning,decorations.pathmorphing}
\usepackage{amsthm}

\usepackage{epstopdf}
\newcommand{\bea}{\begin{aligned}}
\newcommand{\eea}{\end{aligned}}
\newcommand{\beq}{\begin{equation}}
\newcommand{\eeq}{\end{equation}}

\newcommand{\bse}{\begin{subequations}}
\newcommand{\ese}{\end{subequations}}

\usepackage{hyperref}
\hypersetup{
     colorlinks   = true,
     citecolor    = red,
     linkcolor    = cyan,
     urlcolor     = cyan,
}

\newcommand{\bmm}{\begin{multline}}
\newcommand{\emm}{\end{multline}}

\begin{document}

\begin{center}    
\end{center}
\title{Zero-point length as a topological protection of black hole regularity}
\begin{center}   
\end{center}

\author{Ankit Anand}
\email{anand@iitk.ac.in} 
\affiliation{Department of Physics, Indian Institute of Technology, Kanpur 208016, India.}

\author{Kimet Jusufi}
\email{kimet.jusufi@unite.edu.mk} 
\affiliation{Physics Department, State University of Tetovo,
Ilinden Street nn, 1200, Tetovo, North Macedonia.}

\author{Cosimo Bambi}
\email{bambi@fudan.edu.cn} 
\affiliation{Center for Astronomy and Astrophysics,
Department of Physics, Fudan University, Shanghai 200438, China}
\affiliation{School of Humanities and Natural Sciences, New Uzbekistan University, Tashkent 100001, Uzbekistan}

\begin{abstract}
We investigate the thermodynamic topology of regular black holes with zero-point length using an extended first law that includes the zero-point length stored in the geometry. By treating the regularization scale $l_0$ as a thermodynamic variable, we analyze the Hessian geometry of the thermodynamic manifold and demonstrate that the vector field $\vec{\phi} = (T, \Psi)$, where $T$ is the temperature and $\Psi$ is the conjugate to $l_0$, never vanishes in the physical parameter space for $l_0 > 0$. This implies the absence of Morse critical points and a vanishing winding number ($W = 0$), indicating topological protection against the formation of naked singularities. Crucially, we show that in the singular limit $l_0 \to 0$, a non-zero winding number ($W = 1$) emerges, characterizing the Schwarzschild singularity as a topological defect. The conservation of this topological invariant under smooth evolution provides a rigorous topological formulation of the weak cosmic censorship conjecture: the presence of zero-point length not only regularizes the spacetime background but also enforces topological protection against the formation of singularities, preventing black hole-to-naked singularity transitions. 
\end{abstract}

\maketitle

\section{Introduction}

The deep connection between black hole thermodynamics and spacetime singularities has been a central theme in gravitational physics since the seminal work of Bardeen, Carter, and Hawking \cite{Bardeen:1973gs}. The weak cosmic censorship conjecture (WCCC) \cite{Penrose:1969pc} posits that gravitational collapse always produces black holes rather than naked singularities, but its general proof remains elusive. Extremal black holes occupy a special position in gravitational physics: they possess a regular event horizon with vanishing surface gravity and hence zero Hawking temperature, while saturating bounds that relate mass, charge, and angular momentum~\cite{Gibbons:1982fy, Hawking:1994ii}.  Approaching the extremal limit represents a singular point of black hole thermodynamics, where the standard thermal description becomes ill-defined or breaks down altogether~\cite{Horowitz:1996fn, Rothman:2000mm, Liberati:2000sq}.  Exceeding extremality leads to the appearance of naked singularities, violating the cosmic censorship conjecture~\cite{Penrose:1969pc, Israel:1986gqz}. This close correspondence between thermodynamic degeneracy and the onset of spacetime pathologies suggests a deep and intrinsic connection between black hole thermodynamics and cosmic censorship.  

Recently, a topological formulation of black hole thermodynamics has been developed, in which thermodynamic equilibrium states are identified with topological defects in a two-dimensional parameter space spanned by intensive and extensive variables \cite{Wei:2021vdx, Wei:2022dzw, Wu:2024txe}. In this construction, one introduces a thermodynamic vector field whose zeros correspond to equilibrium configurations, while the associated winding number defines a topological invariant that classifies different equilibrium branches and phase transitions. Crucially, this invariant is independent of local thermodynamic response functions such as heat capacity or compressibility, thereby providing a model-independent probe of the global structure of black hole phase space \cite{Wu:2024txe, Wu:2024rmv, Wu:2024asq}.

Regular black holes, which avoid 
singularities \cite{Bardeen:1968, Hayward:2005gi, Ayon-Beato:1998hmi}, have attracted considerable attention. A common feature of such models is the introduction of a magnetic or electric charge,  which can make the spacetime unstable due to the Schwinger effect.   In the present work, we employ the neutral case of the string T-duality–inspired metric introduced in Ref. \cite{Nicolini:2019irw}, which depends on a single parameter, the string tension. The interested reader can also see the  charged black hole in T-duality \cite{Gaete:2022ukm}. The main parameter is the zero-point length given by $l_0 = 2\pi\sqrt{\alpha'}$, where $\alpha'$ is the Regge slope. This scale often represents a zero-point length \cite{Padmanabhan:1996ap} or the minimal measurable distance \cite{Garay:1994en}, emerging from considerations in quantum gravity. This metric describes a neutral, static spacetime and therefore does not suffer from the Schwinger instability that affects regular black-hole metrics obtained by coupling gravity to non-Maxwell electrodynamics. Using this metric, the geodesic completeness of the spacetime was recently demonstrated in Ref. \cite{Jusufi:2024dtr}, work related to bouncing cosmology \cite{Jusufi:2022mir}, as well as ER=EPR and wave function collapse \cite{Jusufi:2025rlr,Jusufi:2025lrq}.

Furthermore, this topological construction admits a natural interpretation in terms of Morse theory. By introducing a smooth thermodynamic potential on the state space, the equilibrium configurations are identified with its critical points, which coincide with the zeros of the thermodynamic vector field. The local winding number associated with each zero is then equal to the Morse index of the corresponding critical point, providing a rigorous mathematical correspondence between thermodynamic stability, phase transitions, and topological charge \cite{MilnorMorse, Arnold, Wei:2022dzw, Wu:2024asq}. In this way, the topological protection of black hole horizons is rooted in the global structure of the thermodynamic potential rather than in perturbative stability criteria.

In this letter, we demonstrate that treating $l_0$ as a thermodynamic variable reveals a profound topological structure protecting black hole regularity. Using Morse theory applied to the thermodynamic potential, we show that the extremal limit does \emph{not} correspond to a critical point when $l_0 > 0$, preventing topological transitions to naked singularities. The phase angle $\theta = \arg(T + i\Psi)$ remains globally well-defined, and the winding number vanishes identically. Also, in the singular limit $l_0 \to 0$, the topological protection breaks down, allowing the Schwarzschild singularity to emerge.

The rest of the letter is structured as follows. In Sec.~\ref{sec:morse}, we develop the Morse theoretic formulation of black hole thermodynamics and set the stage for computation. In Sec.~\ref{Sec:Regular black hole with zero point length} we revisit the regular black hole and its thermodynamics. In Sec.~\ref{Sec:Topological charge Computation}, we compute the topological charge for this regular black hole for both cases, i.e., the non-extremal and the extremal cases. In Sec.~\ref{Sec:Singular Limit and Topological Breakdown}, we compute the topological charge for the singular limit $l_0 \to 0$ and show the topological breakdown. Finally, we end with the conclusion and discussion in Sec.~\ref{Sec:Conclusion and Discussion}.

\section{Morse-theoretic formulation of black hole thermodynamics}
\label{sec:morse}

In this section, we provide a rigorous Morse-theoretic interpretation of the thermodynamic-topological structure underlying black hole equilibrium states~\cite{MilnorMorse, GuilleminPollack}. This framework clarifies the mathematical origin of the topological protection mechanism discussed in the previous sections and establishes its generality, independent of the specific black hole solution.

\subsection{Thermodynamic potential and critical points}

Let $\mathcal{M}$ denote the thermodynamic parameter space of a black hole, coordinatised by a pair of extensive variables $(S,X)$, where $S$ is the entropy and $X$ represents a conserved quantity such as electric charge, angular momentum, or pressure-related variables in extended phase space~\cite{Kastor:2009wy,Dolan:2010ha}. We assume the existence of a smooth thermodynamic potential as follows
\begin{equation}
\mathcal{F} : \mathcal{M} \rightarrow \mathbb{R} \ ,
\end{equation}
which may be identified with the ADM mass $M(S,X)$ or with an appropriate Legendre-transformed free energy, depending on the ensemble~\cite{Hawking:1975vcx, Chamblin:1999tk}.

The equilibrium configurations are determined by the stationarity condition,
\begin{equation}
\nabla \mathcal{F} = 0 \ ,
\label{crit_cond}
\end{equation}
which explicitly reads
\begin{equation}
\partial_S \mathcal{F} = T \ , \qquad
\partial_X \mathcal{F} = \mu_X \ ,
\end{equation}
where $T$ is the Hawking temperature, and $\mu_X$ is the thermodynamic conjugate to $X$~\cite{Bardeen:1973gs}. The extremal black holes correspond to degenerate equilibrium points for which the temperature vanishes and thus appear as critical points of $\mathcal{F}$ in the thermodynamic manifold $\mathcal{M}$~\cite{Hawking:1994ii, Garfinkle:1993xk}.

Near an isolated critical point $(S_i,X_i)$, the local structure of $\mathcal{F}$ is governed by its Hessian matrix
\begin{equation}
H_{ab} = \frac{\partial^2 \mathcal{F}}{\partial X^a \partial X^b} ,
\qquad a,b = 1,\dots,\dim\mathcal{M} \ .
\label{general_hessian}
\end{equation}
The Morse index $\mu_i$ is defined as the number of negative eigenvalues of $H_{ab}$ evaluated at the critical point~\cite{MilnorMorse}. This index characterizes the local topology of the thermodynamic potential near extremality.

We now define the thermodynamic vector field
\begin{equation}
\vec{\phi} = \left( \phi_1 , \phi_2 \right) = \nabla \mathcal{F} = \left( \partial_S \mathcal{F},\partial_X \mathcal{F} \right) \ ,
\label{phi_def}
\end{equation}
which vanishes precisely at the critical points of $\mathcal{F}$. Away from these points, one may introduce a phase angle
\begin{equation}
\theta = \arg(\phi_1 + i \phi_2) \ ,
\end{equation}
defining a continuous map
\begin{equation}
\theta : \mathcal{M}\setminus\{\vec{\phi}=0\} \rightarrow S^1 \ .
\end{equation}


Let $\mathcal{F}=\mathcal{F}(X^a)$ be a smooth thermodynamic potential defined on the thermodynamic manifold $\mathcal{M}$, where
\begin{equation}
X^a = (S, X_1, X_2, \dots)
\end{equation}
denote extensive thermodynamic variables such as entropy, charge, angular momentum, or thermodynamic volume.

For a point $X^a=X_i^a$, a Morse critical point exists if
\begin{equation}
\partial_a \mathcal{F}\big|_{X_i}=0 \ .
\end{equation}
The Morse index $\mu_i$ is then given by the number of negative eigenvalues of $H_{ab}$, the associated Hessian matrix defined in Eq.~\eqref{general_hessian}, evaluated at $X_i$.


Let $C \subset \mathcal{M}$ be a closed contour that does not intersect any critical point. The winding number associated with $\vec{\phi}$ is defined as
\begin{equation}
W = \frac{1}{2\pi} \oint_C d\theta \ .
\label{winding_def}
\end{equation}
This quantity is a homotopy invariant of the map $\theta$ and is therefore unchanged under smooth deformations of the contour or the thermodynamic parameters~\cite{GuilleminPollack, Nakahara}.


Since the winding number \eqref{winding_def} is conserved under smooth thermodynamic evolution, any continuous path in parameter space is forbidden from crossing an extremal boundary unless the thermodynamic description itself breaks down. Consequently, configurations lying beyond extremality—identified with naked singularities—belong to thermodynamically disconnected sectors. This establishes a Morse-theoretic formulation of the weak cosmic censorship conjecture: extremal black holes act as topologically protected critical points, and the formation of naked singularities is obstructed by the conservation of a global topological invariant~\cite{Penrose:1969pc, Israel:1986gqz}.

\section{Regular black hole with zero point length}\label{Sec:Regular black hole with zero point length}

It has been shown that the regularized position-space propagator with a zero-point length emerges from closed string theory in a spacetime with a compactified dimension \cite{Smailagic:2003hm,Fontanini:2005ik}. The key mechanism is T-duality, which relates large and small scales and introduces a fundamental length $l_0$, acting as a UV regulator. The setup considers a closed string in a 4+1 dimensional spacetime, where the fifth dimension $x^5$ is compactified on a circle of circumference $l_0 = 2\pi R$. For closed strings in a compact dimension, the mass spectrum is characterized by two quantum numbers:
   $n $, the Kaluza-Klein excitation level (discrete momentum quantum number), and $w$, the string winding number around the compact dimension. The string mass spectrum for a single compact dimension of radius $R$ is given by \cite{Smailagic:2003hm,Fontanini:2005ik}:
\begin{equation}
m^2 = \frac{1}{2\alpha'} \left( n^2 \frac{\alpha'}{R^2} + w^2 \frac{R^2}{\alpha'} \right) + \cdots
\label{eq:mass_spectrum}
\end{equation}
where $\alpha'$ is the string tension. This spectrum has the fundamental T-duality symmetry: under the transformation $R \leftrightarrow \alpha'/R$ with simultaneous exchange $n \leftrightarrow w$, the spectrum remains invariant. The invariant length scale under this duality is $\sqrt{\alpha'}$, which becomes a candidate for the fundamental \textit{zero-point length} of spacetime. To incorporate the effect of zero-point length we first review the quantum-corrected static interaction potential derived from field theory with path integral duality~\cite{Padmanabhan:1996ap}. The momentum-space massive propagator induced by path integral has been obtained in string T-duality \cite{Smailagic:2003hm}
\begin{equation}
G(k) = -\frac{l_0}{\sqrt{k^2+m_0^2}}\, K_1\!\left(l_0 \sqrt{k^2+m_0^2}\right)  \ ,
\end{equation}
where $l_0$ is the zero-point length with $l_0 = 2\pi\sqrt{\alpha'}$, where $\alpha'$ is the Regge slope.  In addition, $K_1(x)$ is a modified Bessel function of the second kind. For the massless case ($m_0=0$), we recover the conventional propagator $G(k) = -k^{-2}$ at small momenta, while at large momenta, exponential suppression cures UV divergences.

For a static external source consisting of two point masses $m$ and $M$ at relative distance $\vec{r}$, the potential is \cite{Nicolini:2019irw, Nicolini:2022rlz}
\begin{equation*}
V_G(r) = -M \int\!\frac{d^3 k \; e^{i\vec{k}\cdot\vec{r}} }{(2\pi)^3}\, G_F(k)\Big|_{k^0=0}\,  = -\frac{M}{\sqrt{r^2 + l_0^2}} \ .
\end{equation*}

Using Poisson's equation gives an energy density function for the bare matter as
\cite{Nicolini:2019irw}
\begin{equation}
\rho^{\rm bare}(r)= \frac{1}{4\pi}\nabla^2 V_G(r)=\frac{3 l_0^2 M}{4 \pi \left( r^2+l_0^2\right)^{5/2}}.\label{density}
\end{equation}
Imposing spherical symmetry in the metric of the form
\begin{equation}
ds^2 = -f(r)dt^2 + f(r)^{-1}dr^2 + r^2 d\Omega^2 \ ,
\end{equation}
with the standard ansatz for the metric function
\begin{equation}
 f(r)=-g_{tt} = g_{rr}^{-1}=1-\frac{2m(r)}{r},
\end{equation}
the mass profile
\begin{equation}
 m(r)=4 \pi \int_0^r \rho^{\rm bare}(r')  r'^2 dr',
\end{equation}
and by using the energy-momentum tensor conservation, the Bardeen-type quantum-corrected regular black hole metric is obtained \cite{Nicolini:2019irw, Nicolini:2022rlz}
\begin{equation}
f(r) = 1 - \frac{2Mr^2}{\sqrt{(r^2 + l_0^2)^3}} \ ,
\label{eq:metric}
\end{equation}
where instead of magnetic charge $g$ we have the fundamental length scale, i.e.  $ g \to l_0$. At short distances $r \ll l_0$ it has a de Sitter core with a repulsive gravity effect, i.e., $f(r) =1-\Lambda r^2/3$, where $\Lambda=6 M/l_0^3$. This repulsive nature of gravity at short scales plays a crucial role in the geodesic completeness of the spacetime as was recently demonstrated in Ref. \cite{Jusufi:2024dtr}. Finally, at large distances $r \gg l_0$ it reproduces the Schwarzschild spacetime. Depending on the mass parameter, we have three regions: \\

i) A supermassive black hole with $M \gg 3 \sqrt{3} l_0/4 $ with outer and inner horizons given by $r_+$ and $r_-$; \\

ii) An extremal black hole configuration with the extremal mass $M=M_{\rm ext}=3 \sqrt{3} l_0/4$ with degenerate horizons $r_-=r_+=r_{\rm ext}=\sqrt{2} l_0$;\\

iii) A particle sector $M \ll 3 \sqrt{3} l_0/4$ with no horizons. \\

This metric has the advantage of being stable and does not suffer from Schwinger instability. The above metric therefore leads to the possibility of stable black hole remnants of Planck mass which, on the phenomenological level, can be a good candidate for dark matter \cite{Jusufi:2025qgd,Carr:2025auw}.

\subsection{Thermodynamic Framework}
In quantum gravity, the minimal length $l_0$ is expected to fluctuate around the Planck scale due to spacetime quantum uncertainty. The amplitude of these fluctuations likely depends on the system's mass scale $M$, being maximal when $M \sim M_{\text{Pl}}$ and suppressed for $M \gg M_{\text{Pl}}$ (supermassive black hole sector) and $M \ll M_{\text{Pl}}$ (particle sector) by a simple power law in mass. In our analysis, we model this by allowing $l_0$ to vary in a small interval $l_0 \in [\ell_{Pl} - \delta l_0(M), \ell_{Pl} + \delta l_0(M)]$, where $\delta l_0(M)/\ell_P \sim M_{\text{Pl}}/M$. Specifically, we model the fluctuation amplitude $\delta l_0(M)$ as mass-dependent:
\begin{equation}
\frac{\delta l_0(M)}{\ell_{Pl}} \sim 
\begin{cases}
\displaystyle\frac{M}{M_{\text{Pl}}}, & M \ll M_{\text{Pl}} \\[10pt]
\displaystyle\frac{M_{\text{Pl}}}{M}, & M \gg M_{\text{Pl}}
\end{cases}
\end{equation}
so that fluctuations are maximal at $M \sim M_{\text{Pl}}$ and suppressed for both very small and very large masses (see also \cite{Jusufi:2025qgd}). This captures the mass-dependent quantum uncertainty in the geometry's minimal scale. Treating $l_0$ as a thermodynamic variable, the mass is a function of entropy $S$ and $l_0$, the extended first law then can be written as
\begin{equation}
dM = T\,dS + \Psi\,dl_0 \ ,
\label{firstlaw}
\end{equation}
where $T = (\partial M/\partial S)_{l_0}$ is the temperature, and $\Psi = (\partial M/\partial l_0)_S$ is conjugate to $l_0$, respectively.  The horizon radius $r_+$ satisfies $f(r_+) = 0$, and gives the mass, and with the help of this and using the metric function~\eqref{eq:metric}, we can compute the temperature. The expression connecting these quantities is
\begin{equation}
M(r_+, l_0) = \frac{\sqrt{(r_+^2 + l_0^2)^3}}{2r_+^2}  \ .
\label{MandT}
 \end{equation}
With the help of the metric function and mass expression in Eq.~\eqref{MandT}, we can compute the expressions for the temperature and $l_0$'s conjugates $\Psi$ as
\begin{eqnarray}\label{Temp_rel}
    T(r_+, l_0) &=& \frac{r_+^2 - 2l_0^2}{4\pi r_+(r_+^2 + l_0^2)} \\ \label{Psi_expr}
    \Psi(r_+, l_0) &=&  \frac{3 l_0 \sqrt{r_+^2 + l_0^2}  }{2 r_+^2}
\end{eqnarray}
The temperature vanishes at extremality: $r_+^2 = 2l_0^2$. From $dM = T\,dS$ (with $l_0$ fixed during integration), by means of the relation
\begin{equation}
S=\int \frac{1}{T}\frac{\partial M}{\partial r_+}dr_+,
\end{equation}
we obtain the expression of entropy as
\begin{align} \notag
S(r_+, l_0)& = \pi (r_+^2 - 2l_0^2) \frac{\sqrt{r_+^2 + l_0^2}}{r_+} \\
& + 3\pi l_0^2 \ln\!\left(r_+ + \sqrt{r_+^2 + l_0^2}\right)+C,
\end{align}
where $C$ is an integration constant.  Choosing  the following form $C=-3\pi l_0^2 \ln \mathcal{C} $, we can rewrite this relation in terms of another constant 
\begin{align} \notag
S(r_+, l_0) &= \pi (r_+^2 - 2l_0^2) \frac{\sqrt{r_+^2 + l_0^2}}{r_+} \\
            &+ 3\pi l_0^2 \ln\!\left( \frac{r_+ + \sqrt{r_+^2 + l_0^2}}{\mathcal{C}} \right).
\end{align}

In order to see the physical meaning of such a constant more clearly, we can write the entropy in terms of the surface area $A=4  \pi r_+^2$ yielding 
\begin{align}\label{Entropy_final}
S &= \frac{A}{4}\left(1-\frac{8\pi \ell_0^2}{A}\right)
\sqrt{1+\frac{4\pi \ell_0^2}{A}} \notag \\
&\qquad+ 3\pi \ell_0^2 \ln\!\left[ \left(\sqrt{\frac{A}{A_0}}+\sqrt{\frac{A+4\pi \ell_0^2}{A_0}}\right)\right] \ ,
\end{align}
where $A_0=4 \pi \mathcal{C}^2$. From the last equation, if we consider a series expansion around $l_0$, we get the expected logarithmic corrections in entropy. 

The constant $\mathcal{C}$ is not yet fixed, and we can analyze two simple choices: First, we can set it to $\mathcal{C} = l_0$, which yields a positive extremal entropy 
\begin{equation}
S_{\text{extr}} = 3\pi l_0^2 \ln(\sqrt{2} + \sqrt{3}).
\end{equation}
However, a more suitable choice can be $\mathcal{C} = \sqrt{2} l_0$, such that we get the minimal surface $A_0=8 \pi l_0^2$,  which in this case yields again a positive extremal entropy
\begin{equation}
S_{\text{extr}} = 3\pi l_0^2 \ln \left(1 + \sqrt{\frac{3}{2}} \right).
\end{equation}

Let us also mention that in principle it is also possible to set $\mathcal{C} = (\sqrt{2}+\sqrt{3}) 2 l_0$, to get a vanishing extremal entropy, i.e., $S_{\text{extr}} =0$. In the present work, we will consider a positive extremal entropy. 

In the limit $r_+ \gg l_0$, corresponding to the supermassive black hole regime with $r_+ \sim 2M \gg \sqrt{2}\, l_0$, the entropy approaches the area law, $S \simeq \pi r_+^2$. The mass function expressed in terms of entropy admits the leading-order expansion
\begin{equation}\label{M_upto_first_order}
M(S,l_0)= \frac{1}{2}\sqrt{\frac{S}{\pi}}+\frac{3\sqrt{\pi}}{4\sqrt{S}}\, l_0^2 + \mathcal{O}(l_0^4) \ ,
\end{equation}
which in the limit $l_0 \to 0$ gives  
\begin{equation}\label{l_0_mass_case}
M(S,l_0) = \frac{1}{2}\sqrt{\frac{S}{\pi}},
\end{equation}
as expected. 


\section{Topological charge Computation}\label{Sec:Topological charge Computation}

In this section, we compute the topological charge by computing the winding number in Eq.~\eqref{winding_def}. The prescription we are going to follow is: we start by defining the thermodynamic vector field $\vec{\phi}$ as
\begin{equation}
\vec{\phi} \equiv \nabla M = (T, \Psi),
\label{vectorfield}
\end{equation}
The general expression for $T$ and $\Psi$ were already derived in Eq.~\eqref{Temp_rel} and Eq.~\eqref{Psi_expr}. For $\vec{\phi} \neq 0$ everywhere on $\mathcal{M}$, the phase angle
\begin{equation}
\theta = \arg(T + i\Psi) \ ,
\end{equation}
is smooth and globally well-defined. Explicitly,
\begin{equation}\label{Theta_def}
\theta  = \arctan\!\left( \frac{\Psi}{T} \right) \ .
\end{equation}

We consider the non-extremal and extremal cases separately.

\subsection{Non-extremal case}

For the non-extremal case, it is seen from Eq.~\eqref{Temp_rel} that we must have $r_+ \neq \sqrt{2} l_0$.  For the non-extremal case, the physical parameter space $\mathcal{M}$ has the following nonzero values
\begin{equation}
T > 0, \qquad \Psi < \infty \quad \Rightarrow \quad \vec{\phi} \neq (0,0) \ .
\end{equation}
Hence, the thermodynamic vector field never vanishes on $\mathcal{M}$. As a consequence, the mass function $M(S,l_0)$ admits no Morse critical points in $\mathcal{M}$. In particular, there is no extremal limit in the non-extremal sector.

Using the thermodynamic vector field in Eq.~\eqref{vectorfield} and Eq.~\eqref{winding_def}, we compute the topological charge on $\mathcal{C}$ as
\begin{align}
W(\mathcal{C}) = \frac{1}{2\pi} \oint_{\mathcal{C}} d\theta =  \frac{1}{2\pi} \oint_{\mathcal{C}} \frac{T d\Psi - \Psi dT}{T^2 + \Psi^2} \ .
\label{eq:winding_integral}
\end{align}
Since $\vec{\phi}$ has a nonzero value in $\mathcal{M}$, the integrand is smooth and $W(\mathcal{C}) = 0$. 

To strengthen the claim, let us sketch the proof of the vanishing winding number in the non-extremal sector. We consider the thermodynamic parameter space
\begin{equation}
\mathcal{M} = \left\{(r_+, l_0) \;\middle|\; r_+ > \sqrt{2}\, l_0 > 0 \right\} \,
\end{equation}
corresponding to non-extremal regular black holes. Let $\mathcal{C} = \partial R$ be a rectangular loop in the $(r_+, l_0)$-plane,  with $R = [a,b] \times [c,d], a > \sqrt{2}\, d$, so that $R \subset \mathcal{M}$. Now, writing
\begin{equation}
dT= \partial_{r_+} T\,dr_+ +\partial_{l_0} T \,dl_0,
\end{equation}
and 
\begin{equation}
d \Psi= \partial_{r_+}  \Psi\,dr_+ +\partial_{l_0}  \Psi \,dl_0,
\end{equation}
one can easily obtain the numerator of Eq.~\eqref{eq:winding_integral} as 
\begin{eqnarray}\label{Numeratio_of_winding_integral}
  T d\Psi - \Psi dT= A(r_+,l_0) dr_++B(r_+,l_0)dl_0,
\end{eqnarray}
where we have defined
\begin{align}
 A(r_+,l_0) &= T\frac{\partial \Psi}{\partial r_+}
    - \Psi\frac{\partial T}{\partial r_+},\\
    B(r_+,l_0) &= T\frac{\partial \Psi}{\partial l_0}
    - \Psi\frac{\partial T}{\partial l_0} .
\end{align}
Putting this into Eq.~\eqref{eq:winding_integral}, we have 
\begin{equation}\label{eq:winding_integral_f}
d\theta:=P dr_+ + Q dl_0 \ ,
\end{equation}
where 
\begin{equation}
P(r_+,l_0)=\frac{A}{T^2+\Psi^2},
\end{equation}
and 
\begin{equation}
Q(r_+,l_0)=\frac{B}{T^2+\Psi^2}\ .
\end{equation}

Our aim is to compute the integral of Eq.~\eqref{eq:winding_integral_f}. By applying Green’s theorem, the integral reduces to
\begin{equation}
\oint_{\mathcal{C}} d\theta = \iint_R \left( \frac{\partial Q}{\partial r_+} - \frac{\partial P}{\partial l_0} \right) dr_+\,dl_0 \ .
\end{equation}
It is easy to verify by a direct, albeit lengthy, computation that
\begin{equation}
\frac{\partial Q}{\partial r_+} = \frac{\partial P}{\partial l_0}.
\end{equation}
This shows that the integrand vanishes identically. Since any closed curve in $\mathcal{M}$ can be approximated by a finite union of such rectangular loops, it follows from Eq.~\eqref{eq:winding_integral} that
\begin{equation}
W(\mathcal{C}) = 0 \ ,\qquad \text{for all closed \;\;\;} \mathcal{C} \subset \mathcal{M} \ .
\end{equation}

The vector field $\vec{\phi}=(T,\Psi)$ lies entirely in the first quadrant of the $(T,\Psi)$-plane. The phase $\theta$ is globally well-defined and admits no branch cuts. The thermodynamic vector field carries zero topological charge in the non-extremal sector. This vanishing winding number indicates topological triviality: the black hole phase space is simply connected with no vortices or topological defects. The regularization scale $l_0$ plays a dual role: it ensures curvature regularity at the origin and controls the topology of the thermodynamic phase space.  

Let us now show the vanishing of the topological charge density. First, we need to define the 2D topological current $ j^a $ ($ a = r_+, l_0 $) as
\[
j^a = \frac{1}{2\pi} \, \epsilon^{ab} \, \partial_b \theta.
\]
Explicitly, we can write the components as
\[
j^{r_+} = \frac{1}{2\pi} Q, \quad j^{l_0} = -\frac{1}{2\pi} P.
\]
The total topological charge density is
\[
\partial_a j^a = \partial_{r_+} j^{r_+} + \partial_{l_0} j^{l_0} 
= \frac{1}{2\pi} \big( \partial_{r_+} Q - \partial_{l_0} P \big).
\]

By Green's theorem in the $(r_+, l_0)$-plane, we saw that for any closed loop $ C \subset \mathcal{M} $,
\[
\frac{1}{2\pi} \oint_C d\theta 
= \frac{1}{2\pi} \iint_R \left( \partial_{r_+} Q - \partial_{l_0} P \right) dr_+ dl_0 = 0,
\tag{8}
\]
where $ R $ is the region enclosed by $ C $. In the region $\mathcal{M}$, $T$ and $\Psi$ are smooth and $T > 0$ (non-extremal).  
Hence, we see the vanishing of the topological charge density
\[
\partial_a j^a = 0 \quad \text{in } \mathcal{M}.
\tag{7}
\]
The extremal case $ T = 0 $ defines a smooth curve in $(r_+, l_0)$ and, as we saw, the field remains finite because $T^2 + \Psi^2 > 0$ everywhere.
In other words, extremal case is a smooth boundary of the thermodynamic parameter space $\mathcal{M}$.  
The winding number $W(C)$ can be nonzero only if $C$ encloses a point where both $T = 0$ and $\Psi = 0$ simultaneously - but here $\Psi \neq 0$. 
Hence, the topological charge density vanishes everywhere in  the physical parameter space, including at extremality, because $\vec{\phi}$ never hits the origin in the $T$-$\Psi$ plane.
In the entire physical thermodynamic parameter space $(r_+ > 0, l_0 > 0)$, the map $(r_+, l_0) \to (T, \Psi)$ is smooth and $\vec{\phi}$ never vanishes.  
Therefore, the topological charge density $\partial_a j^a$ vanishes identically, and the winding number $W(C)$ is zero for all contractible loops.  
The extremal boundary $r_+ = \sqrt{2} l_0$ is a smooth curve in parameter space, not a topological singularity.

\subsection{Extremal case}

At extremality, the horizon radius saturates the minimal value
\begin{equation}
r_+ = \sqrt{2}\, l_0 \,,
\end{equation}
which corresponds to zero temperature $T=0$, while the conjugate potential remains finite and positive
\begin{equation}
\Psi_{\rm ext} = \frac{3\sqrt{3}}{4} \ .
\end{equation}
Hence, the thermodynamic vector field at extremality is
\begin{equation}
\vec{\phi}_{\rm ext} = (T, \Psi) = (0, \Psi_{\rm ext}) \neq (0,0) \ .
\end{equation}

\begin{figure}[ht]
\centering
\includegraphics[scale=0.3]{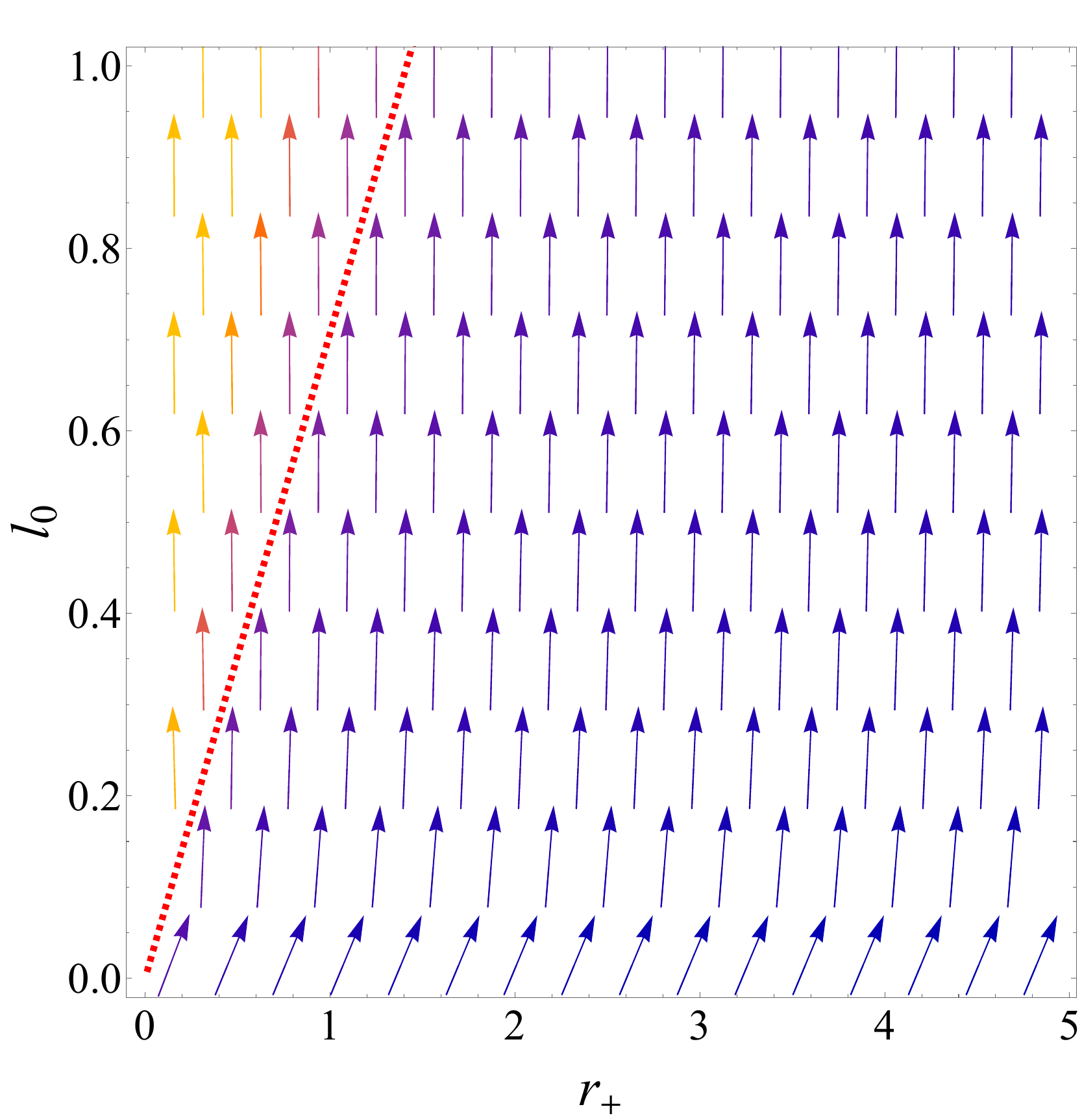}
\caption{{\it{Plot of the vector field $(T,\Psi)$ in the parametric space $(r_+, l_0)$. The region below $l_0<\sqrt{2}/r_+$ represents the case of regular black hole with $l_0>0$ and $r_+>0$. The dashed red line is the extremal black hole case $r_+=\sqrt{2} l_0$.}}}
\label{fig:Equal Case}
\end{figure}

This shows explicitly that extremality does \emph{not} correspond to a zero of $\vec{\phi}$, so it is not a Morse critical point of the mass function $M(S, l_0)$.
The phase angle at extremality is
\begin{equation}
\theta_{\rm ext}  
= \lim_{T \to 0^+} \arctan\left(\frac{\Psi_{\rm ext}}{T}\right) = \frac{\pi}{2} \,,
\end{equation}
which is globally well-defined and continuous. This fact is illustrated in Fig.~\ref{fig:Equal Case}. This apparent singularity is just a coordinate singularity. If we take a closed contour $\mathcal{C}$ in the $(r_+, l_0)$-plane that encloses a segment of the extremal line, $d\theta$ is smooth and the integrand of the winding number formula $W(\mathcal{C})$ is non-singular everywhere on $\mathcal{C}$. Applying Green's theorem as in the non-extremal case, we see again that the winding number around the extremal line vanishes. Extremal black holes do not introduce any topological defect. The vector field $\vec{\phi}$ is nonzero everywhere, the phase $\theta$ is smooth, and the winding number vanishes. Therefore, the extremal configurations represent smooth boundaries of the thermodynamic parameter space rather than topologically nontrivial points; the coordinate $\theta$ becomes singular (a coordinate singularity, not a physical one), and the spacetime geometry remains regular. 

\subsection{Topological characterization of the extremality line}

Let us elaborate further and show that the extremality line $r_+ = \sqrt{2}l_0$ in black hole parameter space, while physically significant as the locus of zero-temperature configurations, does not constitute a topological defect. The proof relies on calculating the winding number of the thermodynamic vector field $(T, \Psi)$ around this line and showing it indeed vanishes. Consider the thermodynamic vector field where the extremality line is defined by $T=0$, i.e., $r_+ = \sqrt{2}l_0$. To analyze the topological properties, we parameterize a small circle of radius $\epsilon$ centered at $(\sqrt{2}a, a)$ on this line:
\begin{equation}
r_+(\varphi) = \sqrt{2}a + \epsilon\cos\varphi,
\end{equation}
\begin{equation}
l_0(\varphi) = a + \epsilon\sin\varphi,
\end{equation}
with $\varphi \in [0, 2\pi]$ and $a > 0$ and $0 < \epsilon \ll a$. Expanding to $\mathcal{O}(\epsilon)$ yields
\begin{equation}
T(\varphi) = \frac{\epsilon(\cos\varphi - \sqrt{2} \sin\varphi)}{6\pi a^2} + \mathcal{O}(\epsilon^2),
\end{equation}
and 
\begin{equation}
\Psi(\varphi) = \frac{3\sqrt{3}}{4} + \mathcal{O}(\epsilon) \equiv \Psi_0 + \mathcal{O}(\epsilon),
\end{equation}
where $\Psi_0 =\Psi_{\rm ext}=3\sqrt{3}/4 > 0$ is constant to leading order. The phase angle then reads 
\begin{equation}
\theta(\varphi) = \arctan\left(\frac{9 \sqrt{3}\, \pi\, a^2}{2\epsilon(\cos\varphi - \sqrt{2}\sin\varphi)}\right) + \mathcal{O}(\epsilon).
\end{equation}
Defining $f(\varphi) = \cos\varphi - \sqrt{2}\sin\varphi$, we find its zeros at
\begin{equation}
f(\varphi)=0 \;\Rightarrow\; \tan\varphi = 1/\sqrt{2},
\end{equation}
with solutions $\varphi_1 = \arctan(1/\sqrt{2})$ and $\varphi_2 = \varphi_1 + \pi$ in $[0, 2\pi)$. From the asymptotic form of $\theta(\varphi)$, as $\epsilon \to 0$,
\begin{equation}
\theta(\varphi) \to 
\begin{cases}
\displaystyle +\frac{\pi}{2} & \text{for } \varphi <\varphi_1, \\
\\
\displaystyle -\frac{\pi}{2} & \text{for } \varphi >\varphi_1.
\end{cases}
\end{equation}
and 
\begin{equation}
\theta(\varphi) \to 
\begin{cases}
\displaystyle -\frac{\pi}{2} & \text{for } \varphi <\varphi_2, \\
\\
\displaystyle +\frac{\pi}{2} & \text{for } \varphi >\varphi_2.
\end{cases}
\end{equation}
Consequently, at $\varphi = \varphi_1$ and $\varphi = \varphi_2$, the phase encodes discontinuities $\Delta\theta_{\varphi_1} =-\pi$ and  $\Delta\theta_{\varphi_2} = +\pi.$ \\

The winding number can be computed from the following relation
\begin{align}\notag
W &= \frac{1}{2\pi}\oint d\theta \\
&= \frac{1}{2\pi}\int_{\text{cont.}} \frac{d\theta}{d\varphi}d\varphi + \frac{1}{2\pi}\sum_i \Delta\theta_i^{\text{jumps}}.
\end{align}

For small $\epsilon$, using $\arctan(x) \approx x$,
\begin{equation}
\theta(\varphi) =
\begin{cases}
\displaystyle \frac{\pi}{2} - \frac{2\epsilon}{9 \sqrt{3}\pi a^2 } f(\varphi), & f(\varphi) > 0, \\
\displaystyle -\frac{\pi}{2} - \frac{2\epsilon}{9 \sqrt{3} \pi a^2 } f(\varphi), & f(\varphi) < 0,
\end{cases}
\end{equation}
so that
\begin{equation}
\frac{d\theta}{d\varphi} = -\frac{2\epsilon}{9 \sqrt{3}\pi a^2 } f'(\varphi).
\end{equation}
The continuous part of the integral therefore vanishes since $f(2\pi)=f(0)=1$. The jump contributions cancel exactly:
\begin{equation}
\sum_{i} \Delta\theta_i^{\text{jumps}} = (-\pi) + (+\pi) = 0.
\end{equation}
Hence,
\begin{equation}
\oint d\theta = 0 \quad \Longrightarrow \quad W = 0.
\end{equation}

Geometrically, in the $(T,\Psi)$-plane, the curve $(T(\varphi),\Psi(\varphi))$ lies entirely in the upper half-plane ($\Psi \approx \Psi_0 > 0$) and does not enclose the origin. The extremality line $r_+ = \sqrt{2}l_0$ carries zero topological charge ($W=0$) and therefore is \textit{not} a topological defect. Only the point $(r_+,l_0)=(0,0)$, where both $T$ and $\Psi$ vanish simultaneously, give rise to a topological defect in this parameter space.

\section{Singular Limit $l_0 \to 0$ and Topological Breakdown}\label{Sec:Singular Limit and Topological Breakdown}

We can now discuss the singular limit $l_0 \to 0$. Since the limit is singular, it cannot be taken continuously within the thermodynamic phase space. In this limit, the mass reduces to Eq.~\eqref{l_0_mass_case} and no independent conjugate variable $\Psi$ exists, or, in other words, $\Psi$ is undefined in this limit. Thus, the thermodynamic vector field collapses from a two-dimensional field to a one-dimensional object. This represents a breakdown of the vector-field topology rather than a smooth limit.

We now establish rigorously that the Schwarzschild singularity, corresponding to the limit $l_0 \to 0$, carries a non-zero topological charge manifested as a unit winding number. This provides a precise topological distinction between regular black holes ($l_0 > 0$) and the singular case ($l_0 = 0$).

In models of regular black holes, such as our T-duality-inspired metric, there exists a fundamental length scale $l_0 > 0$, often related to the non-singular core or a quantum gravity scale. In the limit $l_0 \to 0$, the entropy reduces to the usual formula $S = \pi r_+^2 \,.$ During the Hawking evaporation process, the black hole loses mass, the horizon radius $r_+$ decreases, and eventually the entropy vanishes: $S \to 0 \,. $ This corresponds to a \emph{singular point} in the thermodynamic phase space $(S, l_0) = (0,0) $, which physically represents the complete evaporation of a classical black hole. 
On the other hand, in the regular black hole case, the presence of $l_0$ introduces a minimal horizon radius $r_\text{extr}$ corresponding to an extremal configuration. Consequently, there is a minimal entropy $ S_\text{extr}> 0 \,. $ Therefore, for $l_0 > 0$, the singular point $(S,l_0) = (0,0)$ lies outside the physical domain, since regular black holes cannot have entropy below $S_\text{extr}$. In other words, regular black holes do not evaporate completely, and their thermodynamic phase space is truncated at a nonzero minimal entropy. The case $l_0 \to 0$ reproduces classical black hole physics and the Hawking evaporation endpoint $S = 0$, while $l_0 > 0$ regularizes the singularity and ensures $S \ge S_\text{extr} > 0$. The point $(S, l_0) = (0,0)$ is thermodynamically singular in the classical theory but unphysical for regular black holes.

\subsection{Singularity as a topological defect}
To compute the topological charge of the singular configuration, we treat it as a mathematical defect and we consider the extended phase space $(r_+,l_0)$ in the whole region. First of all, let us mention that the metric function
\begin{equation}
f(r)=1-\frac{2Mr^2}{\left(r^2+l_0^2\right)^{3/2}}
\end{equation}
depends only on $r^2$ and $l_0^2$; it satisfies $f(-r,-l_0)=f(r,l_0)$, implying a $\mathbb{Z}_2$ symmetry under $r\to -r$. This parity symmetry of specific metrics was pointed out in \cite{Zhou:2022yio} and plays a crucial role in the geodesic completness; for example, this means that our Bardeen-type metric has a clear advantage compared to the Hayward metric. The extension to negative $r$ therefore corresponds to a simple antipodal (parity) continuation and is fully analytic across $r=0$, which is neither a physical boundary nor a curvature singularity. Since curvature invariants and the area element depend only on $r^2$, the regions $r>0$ and $r<0$ are geometrically equivalent.

The situation with the horizon parameter $r_+$ is a little bit different. That is, although extending the horizon parameter to $r_+\in\mathbb{R}$ is mathematically justified, it introduces a singular point at $r_+=0$. Therefore, the extension is essential for the topological analysis of the thermodynamic phase space, as it allows the construction of closed contours in the parametric $(r_+,l_0)$-plane encircling the singular point $(r_+,l_0)=(0,0)$: because the winding number depends only on the continuity of the map $(r_+,l_0)\mapsto(T,\Psi)$, and not on the physical interpretation of intermediate points, the antipodal extension provides a consistent framework for defining the associated topological invariant. We therefore consider a small circular contour $\mathcal{C}_\epsilon$ of radius $\epsilon$ enclosing the origin.
\begin{equation}
r_+(\varphi)=\epsilon\cos\varphi, \qquad 
l_0(\varphi)=\epsilon\sin\varphi, \qquad \varphi\in[0,2\pi],
\label{eq:contour_parameterization}
\end{equation}
with $\epsilon\to0^+$. Again, the region $r_+<0$ is physically not well motivated, but this extension is mathematically allowed since the winding number depends only on the continuity of the map $(r_+,l_0)\mapsto(T,\Psi)$. 

Substituting \eqref{eq:contour_parameterization} into the expressions for the temperature and conjugate potential and expanding to leading order in $\epsilon$ yields
\begin{equation}
T(\varphi)=\frac{\cos^2\varphi-2\sin^2\varphi}{4\pi\epsilon\cos\varphi},
\qquad
\Psi(\varphi)=\frac{3\sin\varphi}{2\cos^2\varphi}.
\label{eq:T_Psi_expansion}
\end{equation}

\begin{figure}[ht]
\centering
\includegraphics[scale=0.3]{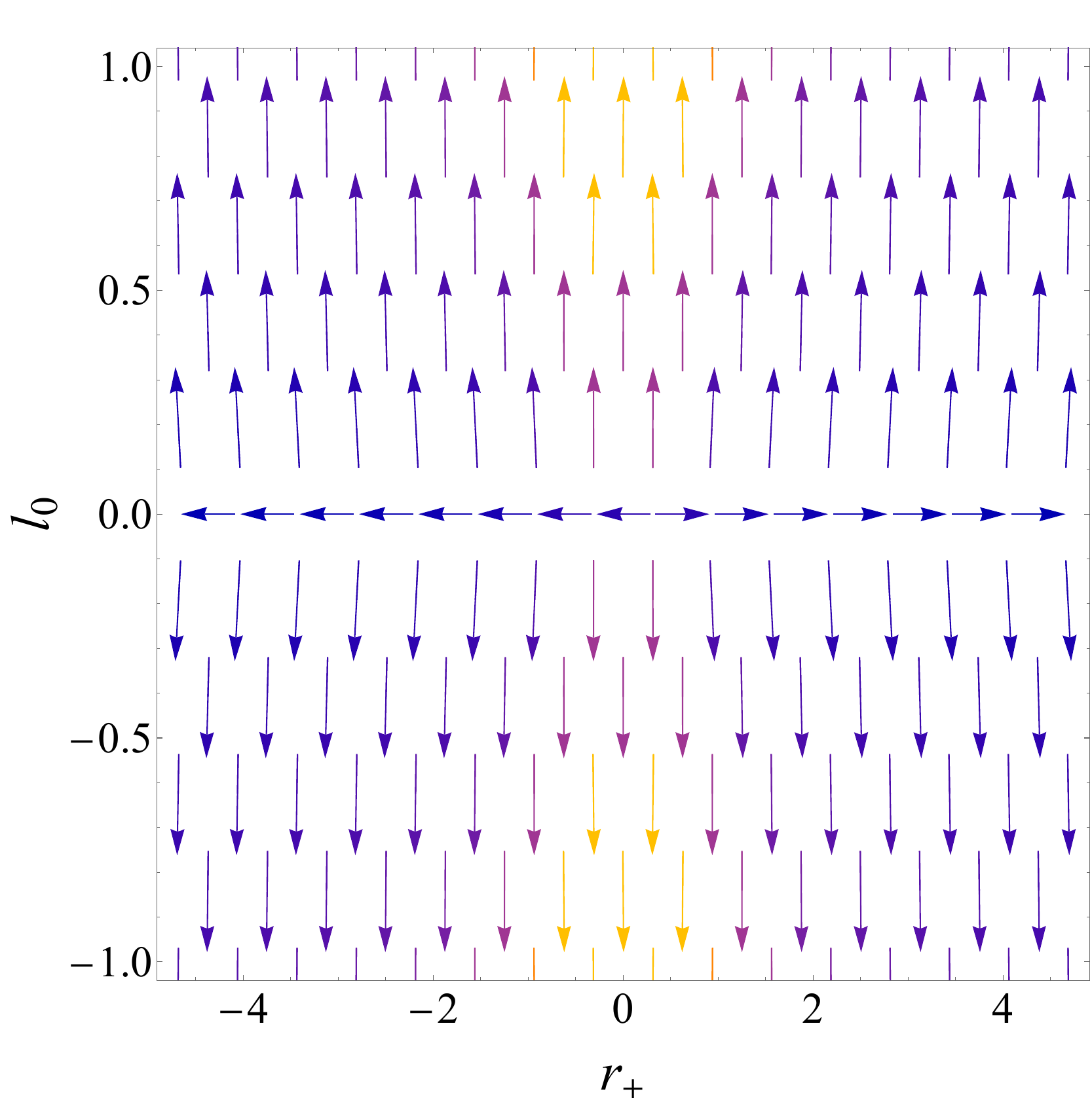}
\caption{{\it{Plot of the vector field $(T, \Psi)$ in the parametric space $(r_+, l_0)$. The region represents the case of positive and negative region for $l_0$ and $r_+$. There is a topological defect in the central region $(r_+=0,l_0=0)$.}}}
\label{fig:positive and negative region}
\end{figure}

The vector field $\vec{\phi}=(T,\Psi)$ becomes singular at $\cos\varphi=0$, i.e., at $\varphi=\pi/2$ and $3\pi/2$. The vector field is illustrated in Fig.~\ref{fig:positive and negative region}. Consider the phase
\begin{equation}
\theta(\varphi)=\arctan\!\left(\frac{\Psi(\varphi)}{T(\varphi)}\right).
\end{equation}
We analyze the behavior of $\theta$ in the neighborhood of
\begin{equation}
\varphi=\frac{\pi}{2}+\eta, \qquad |\eta|\ll 1 .
\end{equation}
Near this point, where $\epsilon>0$ is a small regulator, we get
\begin{equation}
\frac{\Psi}{T}\simeq 3\pi\,\frac{\epsilon}{\eta},
\end{equation}
and the phase takes the approximate form
\begin{equation}
\theta\simeq \arctan\!\left(3\pi\,\frac{\epsilon}{\eta}\right).
\end{equation}
Taking the limits on the two sides of $\eta=0$, we find
\begin{align}
\lim_{\eta\to 0^{\pm}}\theta &= \pm \frac{\pi}{2},
\end{align}
Therefore, when $\varphi$ crosses $\pi/2$, the phase undergoes a
discontinuous jump
\begin{equation}
\Delta\theta_{\pi/2}
=\left(+\frac{\pi}{2}\right)-\left(-\frac{\pi}{2}\right)
=+\pi .
\end{equation}
The geometric interpretation is the following: in the complex $Z$-plane, the real part $T$ changes sign at
$\varphi=\pi/2$, while the imaginary part $\Psi$ remains finite due to
the regulator $\epsilon$. As a result, the trajectory of $Z$ crosses the
imaginary axis and winds halfway around the origin, producing a
topological phase contribution of $\pi$.
An identical jump occurs at $\varphi=3\pi/2$,
\begin{equation}
\Delta\theta_{3\pi/2}
=\left(+\frac{\pi}{2}\right)-\left(-\frac{\pi}{2}\right)
=+\pi .
\end{equation}

The winding number associated with $\mathcal{C}_\epsilon$ is therefore
\begin{equation}
W(\mathcal{C}_\epsilon)=\frac{1}{2\pi}\oint_{\mathcal{C}_\epsilon} d\theta
=\frac{1}{2\pi}\big(\pi+\pi\big)=1.
\end{equation}
Taking the limit $\epsilon\to0$, we obtain $\lim_{\epsilon\to0}W(\mathcal{C}_\epsilon)=1,$
identifying the Schwarzschild singularity as a topological defect with unit winding number. The sign reflects the chosen orientation of the contour.

\section{Topological Conservation and Singularity Exclusion}
\label{sec:topo-conservation}

The vanishing winding number $W=0$ for all $l_0>0$ has direct consequences for the classical evolution of regular black holes.  
Because $W$ is a homotopy invariant, it is conserved under any smooth deformation of the spacetime that preserves regularity.

Consider a one-parameter family of regular black hole solutions $g_{\mu\nu}(\lambda)$ with $l_0(\lambda)>0$, smoothly connected for $\lambda$ in some region. 
The associated thermodynamic vector field $\vec{\phi}=(T,\Psi)$ varies continuously with $\lambda$ and remains nowhere vanishing on the physical parameter space
\begin{equation}
\mathcal{M} = \left\{(r_+, l_0) \;\middle|\; r_+ \geq  \sqrt{2}\, l_0 > 0 \right\} .
\end{equation}
For any closed contour $\mathcal{C}\subset\mathcal{M}$, the winding number $W(\mathcal{C})$ depends only on the homotopy class and therefore remains invariant under smooth evolution.  Since $\vec{\phi}\neq 0$ everywhere for $l_0>0$, one finds $W=0$ for all regular configurations.

In contrast, the singular Schwarzschild limit $l_0\to 0$ corresponds to a breakdown of the thermodynamic description: $\Psi$ diverges and $\vec{\phi}$ becomes ill-defined.  
A contour enclosing the point $(r_+=0,l_0=0)$ acquires a nonzero winding number $ W_{\mathrm{sing}}= 1$, identifying the Schwarzschild singularity as a topological defect. Since the winding number cannot change under smooth evolution unless $\vec{\phi}$ vanishes somewhere along the contour, a regular black hole with $W=0$ cannot evolve continuously into a singular configuration with $W\neq 0$.  
Regular and singular black holes therefore belong to distinct topological sectors of the thermodynamic phase space, separated by a topological barrier.

This establishes a topology-based no-go theorem: \textit{the presence of a nonzero minimal length $l_0$ not only regularizes curvature invariants but also enforces topological protection against the formation of singularities.}

\section{Conclusion and Discussion}\label{Sec:Conclusion and Discussion}

The result of our work establishes a profound connection between quantum-gravity regularization, black hole thermodynamics, and topological conservation laws. The zero-point length $l_0$, originally introduced to cure curvature singularities from a geometric perspective, also plays the role of a \emph{topological regulator}. When $l_0>0$, the thermodynamic phase space is simply connected and topologically trivial, with vanishing winding number $W=0$. In contrast, in the singular limit $l_0\to 0$, a topological defect with unit charge $|W|=1$ emerges, corresponding precisely to the Schwarzschild curvature singularity.

For regular black holes with $l_0>0$, the thermodynamic vector field $\vec{\phi}=(T,\Psi)$ has no zeros anywhere in the physical parameter space $\mathcal{M}=\{(S,l_0): S>S_{\text{extr}}(l_0)\}$. As a consequence, the phase angle associated with $\vec{\phi}$ is globally well defined, and all closed contours in $\mathcal{M}$ carry zero winding number. The thermodynamic phase space is therefore simply connected and free of topological defects. In contrast, when the regulator is removed by taking the limit $l_0\to 0$, the point $(S,l_0)=(0,0)$ becomes a genuine topological defect carrying a nontrivial winding number. This defect coincides with the emergence of the Schwarzschild singularity.

The transition from $W=0$ to $|W|=1$ is necessarily discontinuous and thus represents a topological phase transition. Such a transition cannot be achieved through smooth variations of thermodynamic parameters, reflecting the homotopy invariance of the winding number. Consequently, no continuous thermodynamic evolution can connect the regular sector with $l_0>0$ to the singular sector at $l_0=0$. 

This observation leads naturally to a form of \emph{topological cosmic censorship}. Regular black holes endowed with a nonzero zero-point length are topologically protected from evolving into naked singularities. The formation of a curvature singularity would require a discontinuous change in the topological structure of the thermodynamic phase space, which is forbidden under smooth thermodynamic evolution. This argument forbids the existence of singular spacetimes with $r_+=0$ and $l_0=0$; however, there exists the possibility of including two asymptotic regions by merging two spacetimes that include only the regions $r \ge r_+ = \sqrt{2}\, l_0$. This construction defines a nontrivial geometry, which can be interpreted as a wormhole-like region. Importantly, this spacetime does not contain a singularity and carries a vanishing topological charge. Such a configuration is therefore allowed.

The winding number $W$ therefore serves as a topological order parameter distinguishing regular black holes from singular ones. Its conservation under smooth deformations provides a robust and model-independent mechanism for cosmic censorship, insensitive to the specific details of the underlying dynamics or matter content. Treating the zero-point length $l_0$ as a thermodynamic variable reveals that the associated thermodynamic geometry exhibits intrinsic topological protection: the vector field $\vec{\phi}$ never vanishes in the physical domain, preventing the appearance of naked singularities through continuous evolution.

In summary, regular black holes with a minimal length scale $l_0>0$ are topologically protected against singularity formation. By analyzing the thermodynamic vector field and its associated winding number, we have shown that regularity is preserved by the trivial topology of the phase space, while singular behavior emerges only through a genuine topological phase transition in the limit $l_0\to 0$. These results suggest that quantum gravity, through the introduction of a minimal length scale, may naturally enforce cosmic censorship via topological mechanisms in thermodynamic phase space. Future investigations may extend this framework to rotating regular black holes, anti-de Sitter backgrounds, and higher-dimensional settings.\\


\section*{Acknowledgements}

Ankit Anand is financially supported by the Institute's postdoctoral fellowship at IIT Kanpur.

\bibliographystyle{utphys}
\bibliography{regular}

\end{document}